\documentclass[aps,prb,twocolumn]{revtex4}
\usepackage{bm,graphicx,graphics,amsmath,amssymb,bm,epsfig,color,pifont}
\usepackage{euscript,tabularx,textcomp}
\usepackage[english]{babel}

\begin{document}

\title{High frequency EPR on dilute solutions\\ of the single molecule
  magnet Ni$_4$}

\author{G. de Loubens} \affiliation{Department of Physics, New York
  University, 4 Washington Place, New York, New York 10003, USA}

\author{A. D. Kent} \affiliation{Department of Physics, New York
  University, 4 Washington Place, New York, New York 10003, USA}

\author{V. Krymov}\affiliation{Department of Physiology and
  Biophysics, Albert Einstein College of Medicine, Yeshiva University,
  1300 Morris Park Avenue, Bronx, New York 10461, USA}
\author{G. J. Gerfen}\affiliation{Department of Physiology and
  Biophysics, Albert Einstein College of Medicine, Yeshiva University,
  1300 Morris Park Avenue, Bronx, New York 10461, USA}

\author{C. C. Beedle}\affiliation{Department of Chemistry and
  Biochemistry, University of California San Diego, La Jolla,
  California 92093-0358, USA}
\author{D. N. Hendrickson}\affiliation{Department of Chemistry and
  Biochemistry, University of California San Diego, La Jolla,
  California 92093-0358, USA}

\date{\today}

\begin{abstract}
  Dilute frozen solutions of the single molecule magnet Ni$_4$ ($S=4$)
  have been studied using high frequency D-band (130 GHz) EPR.
  Despite the random orientation of the molecules, well defined EPR
  absorption peaks are observed, due to the strong variation of the
  splittings between the different spin-states on magnetic field.
  Temperature dependent studies above 4~K and comparison with
  simulations enable identification of the spin transitions and
  determination of the Hamiltonian parameters. The latter are found to
  be close to those of Ni$_4$ single crystals. No echo was detected
  from Ni$_4$ in pulsed experiments, which sets an upper bound of
  about 50~ns on the spin coherence time.
\end{abstract}


\maketitle

Single molecules magnets (SMMs) have been suggested as candidates for
qubits in quantum processors \cite{leuenberger01}. However, the
coherence time ($T_2$) of high-spin molecules has not yet been
determined. Quantum tunneling of magnetization (QTM) has been widely
studied in SMMs \cite{friedman96,thomas96,wernsdorfer99}. A recent
focus is on coherent QTM in which the tunneling rates may be faster
than the rate of decoherence \cite{hill03,barco04b,loubens07b}. Due to
inhomogeneous broadening, only a lower bound ($\approx 0.5$~ns on most
of the measured SMMs) of $T_2$ can be extracted. Furthermore, in SMM
crystals the strong dipolar interactions between the molecules
(separated by only 1~nm) is expected to drastically reduce the
coherence time. In order to determine the latter (\emph{e.g.} through
spin echo) and to coherently manipulate the magnetization with
microwave pulses in SMMs, it thus may be necessary to work with dilute
ensembles of molecules.

SMMs may be diluted in solvents. The drawback of such a method is that
the molecules are randomly oriented. However, magnetometry
\cite{novak95} and EPR experiments \cite{barra97} on randomly oriented
powder have been succesfully performed. Only recently have dilute
frozen solutions been studied by high frequency (HF) techniques.
Frequency domain spectroscopy showed that the zero field splitting
(ZFS) parameters of Mn$_{12}$ clusters diluted in a frozen solvent
were almost the same as in the crystal form \cite{hallak07}. Even more
interesting, standard microwave pulse experiments have determined a
long intrinsic coherence time (few microseconds at $T=2$~K) in doped
antiferromagnetic wheels (low-spin) \cite{ardavan07}.

In this paper we present continuous wave (cw) HF EPR experiments
performed on dilute frozen solutions of the SMM Ni$_4$ ($S=4$) as a
function of temperature between 6 and 30~K. Well defined transitions
are observed in the 0--7~Tesla range at $f=130$~GHz. Simulations using
ZFS parameters close to the ones determined from single crystals
\cite{edwards03} and $D$ and $g$ strains qualitatively reproduce the
spectra measured. No spin echo from the Ni$_4$ clusters was observed
in pulse experiments, which sets an upper limit for $T_2$ of about
50~ns.

\begin{figure}
  \includegraphics[width=7cm]{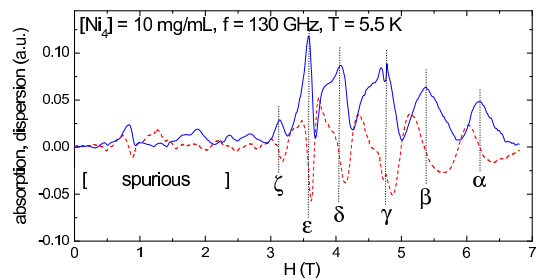}
  \caption{(Color online) EPR spectrum of a 10~mg/mL Ni$_4$ solution
    at $f=130$~GHz and $T=5.5$~K. The dashed and solid lines are
    respectively the absorption and dispersion channels (a
    12~G$_\text{pp}$ field modulation is used). The spurious signal at
    low field was also observed without Ni$_4$ molecules in the 1:1
    mixture of toluene and dichloromethane.}
  \label{spec5p5K}
\end{figure}

[Ni(hmp)(dmb)Cl]$_4$, henceforth referred to as Ni$_4$, is a
particularly clean SMM with no solvate molecules present in its
crystal phase and only 1\% (natural abundance) of nuclear spins on the
transition metal sites \cite{yang03,yang06}. This results in narrower
EPR peaks than in many SMMs \cite{edwards03}. The spin Hamiltonian of
Ni$_4$ is to first approximation:
\begin{equation}
  \mathcal{H}=DS_z^2+BS_z^4+C(S_+^4+S_-^4)+\mu_B\vec{H}.\hat{g}.\vec{S},
\label{ham}
\end{equation}
where the first term is the uniaxial anisotropy, the second and third
terms are $4^{th}$-order anisotropy terms, and the last term is the
Zeeman energy. The $S=4$ ground state of the molecule at low
temperature is the consequence of ferromagnetic exchange interactions
between the four Ni$^{\text{II}}$ ($S=1$) ions. The uniaxial
anisotropy leads to a large energy barrier $|D|S^2\approx12$~K to
magnetization reversal between states of projection $S_z=\pm4$ along
the easy axis of the molecule. The ZFS parameters have been measured
on single crystals using HF EPR \cite{edwards03}. In particular, the
strong $C$--term ($\approx 10$~MHz) explains the fast tunneling
observed at zero field \cite{kirman05}, and a recent analysis shows
that the fourth order terms in Eq.~\ref{ham} may result from a finite
ratio of the (second order) single ion anisotropies to exchange
constant \cite{wilson06}.

To prepare dilute solutions of Ni$_4$, dried crystals were dissolved
in a 1:1 mixture of toluene and dichloromethane. This glass was chosen
due to solubility restraints. $^1$HNMR spectra were collected for
[Ni(hmp)(dmb)Cl]$_4$ and the analogous complexes
[Ni(hmp)(CH$_3$OH)Cl]$_4$ and [Ni(hmp)(CD$_3$OD)Cl]$_4$, as well as
the free ligand 2-hydroxymethylpyridine (hmpH). The proton signals for
the bound hmp$^-$ ligand were positively identified through comparison of
the [Ni(hmp)(CH$_3$OH)Cl]$_4$ and deuterated [Ni(hmp)(CD$_3$OD)Cl]$_4$
complexes. Comparison of the [Ni(hmp)(dmb)Cl]$_4$ and free hmpH
spectra confirmed that no free ligand is present, and thus, the
[Ni(hmp)(dmb)Cl]$_4$ complex is intact in solution. Most of our EPR
experiments were carried out on a 10~mg/mL solution, corresponding to
an average distance of 6~nm between molecules.

D-band (130~GHz) EPR spectra were acquired on a spectrometer capable
of working in cw and pulse modes. The quadrature detection microwave
bridge was designed and built by HF EPR Instruments, Inc. (V. Krymov,
New York). During the acquisition in cw mode, a field modulation of
12~G$_\text{pp}$ is generated at 100~kHz by a coil surrounding the
cavity, and the dispersion and absorption channels \cite{krymov03} are
recorded as a function of the external field, generated by a 7~T
superconducting magnet. Fig.~\ref{spec5p5K} shows the result of such
an experiment performed at 5.5~K on a 10~mg/mL Ni$_4$ solution
contained in a quartz capillary tube (active volume 0.2~$\mu$L). Large
absorption peaks (labeled $\alpha$ to $\zeta$ with decreasing field)
can be observed above 2.5~T. They are attributed to transitions
between spin-states of the dissolved Ni$_4$ molecules. The narrow
signal around 4.6~T ($g\approx 2$) and the spurious absorption at low
field were also measured without Ni$_4$ clusters in the solvent, in
contrast to the high field peaks. These do not affect the high field
part of the spectrum, and we will not discuss them further.

\begin{figure}
  \includegraphics[width=7cm]{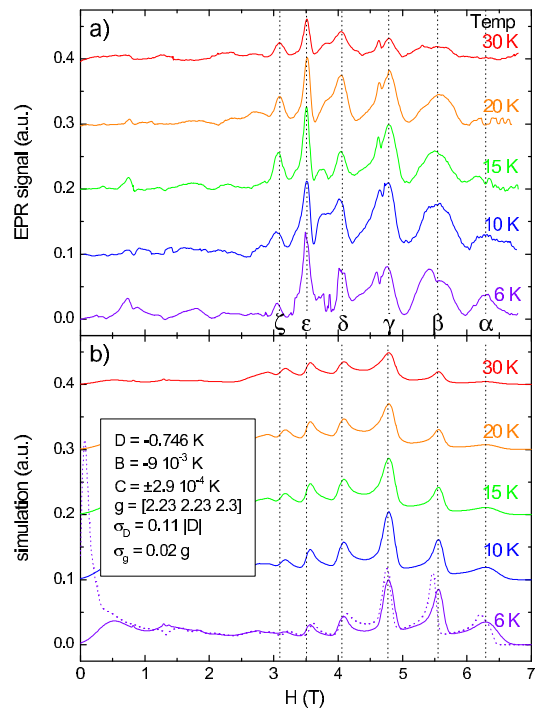}
  \caption{(Color online) a) EPR signal versus temperature. b) The
    dashed curve is the simulation at 6~K using the ZFS parameters of
    the single crystal (see text) and a 0.1~T gaussian broadening. The
    continuous curves have been calculated using the parameters
    indicated in the inset, that include $D$ and $g$ strains.}
  \label{vsT}
\end{figure}

\begin{figure}
  \includegraphics[width=8cm]{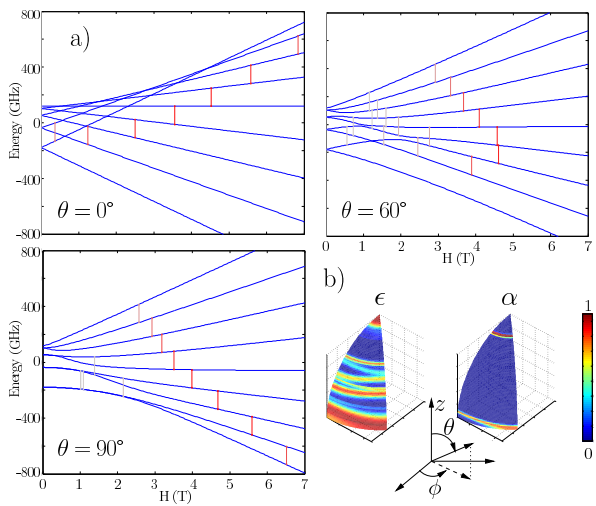}
  \caption{(Color online) a) Energy levels for 3 different angles
    between the applied field and the molecular easy axis. The
    transitions corresponding to $f=130$~GHz are shown. The stronger
    the red color, the larger the matrix element of the transition.
    b) Orientations contributing to transitions $\alpha$ and
    $\epsilon$ (\emph{cf.} Fig.~\ref{spec5p5K} and \ref{vsT}). Note
    that in a) and b), all allowed transitions are displayed without
    regard to the temperature.}
  \label{angle}
\end{figure}

To address the origin of the transitions observed in
Fig.~\ref{spec5p5K}, we discuss the dependence of the EPR signal on
temperature and the comparison with simulations, shown in
Fig.~\ref{vsT}a and b respectively. The latter were performed using
the EasySpin toolbox \cite{easyspin}, which allows calculation of EPR
spectra \cite{stoll06} and provides the visualisation tools used in
Fig.~\ref{angle}. 2\textdegree$\times$2\textdegree~cells in one octant
of the upper hemisphere were used to perform the powder average.
First, we used the ZFS parameters of Ni$_4$ single crystal ($D =
-0.748(5)$~K, $B = -6\times10^{-3}$~K, $C = \pm 2.9\times10^{-4}$~K,
$g_x=g_y=2.23$, $g_z=2.3$) \cite{edwards03, kirman05}, and a
convolution by a gaussian of width 0.1~T (dashed curve on
Fig.~\ref{vsT}b at 6~K). The position of the high field peaks are
slightly offset with respect to the data, and a large absorption peak
close to zero field is calculated, but not detected.

To improve the position of the calculated peaks, we used slightly
different $D$ and $B$ parameters ($D = -0.746(5)$~K, $B =
-9\times10^{-3}$~K), and introduced $D$ and $g$ strains
($\sigma_D=0.11|D|$ and $\sigma_{\hat{g}}=0.02\hat{g}$), which wash
out the low field transitions and lead to an overall better fit to the
spectrum. It is known that $D$ and $g$ strains influence the
lineshapes in SMM single crystals \cite{hill02}. Random dipolar
fields, also expected to play a role, should be less important in
dilute solutions due to the larger average distance between
neighboring molecules. The simulations reproduce rather well the
dependence of the experimental data on temperature. In particular, it
can be seen that, on top of the global decrease of the absorption, the
simulations reproduce the extinction of the peaks $\alpha$ and $\beta$
as the temperature increases. This is expected from Boltzmann
statistics, since the difference in population between the lowest
energy levels decreases exponentially with temperature (note that
$f=130$~GHz corresponds to $T=6.2$~K). Therefore, $\alpha$ is
associated with a transition from the ground state to the first
excited state. However, we were unable to reproduce the lineshapes and
the very intense $\epsilon$ peak. We explored the possibility of
having two Ni$_4$ species (dissolved and precipitated \cite{hallak07})
in the solution without success. We believe that our dilute solutions
are free of precipitate, and it was checked after each experimental
run that the dilute solution contained in the quartz tube was still
clear. Increase of the $D$ and $g$ strains does not improve the fit
either. The values used in the simulations are already larger than the
ones fitted on single crystals \cite{hill02} by a factor 2 to 10.
This probably results from the variation in the molecular environment
of the SMM clusters in the frozen solution. It is known, \emph{e.g.}
in Mn$_{12}$ crystals, that this environment plays a critical role
\cite{barco05}. In solution, the strain could also depart from axially
symmetric. $E$-strain can not be excluded, as simulations performed
with $\sigma_E=0.05$~K slightly broaden peaks $\alpha$ to $\gamma$.

To better understand the nature of the transitions involved in the EPR
spectrum, Fig.~\ref{angle}a shows the energy levels as a function of
the angle $\theta$ between the easy axis of the molecules and the
magnetic field (the angle $\phi$ in the hard plane is set to
0\textdegree). The well-known diagram is obtained for
$\theta=0$\textdegree, with the low lying transitions at low field. In
our experimental case, the ground state transition
($m=-4\rightarrow-3$) at low field is not observed because the ZFS of
Ni$_4$ is about 140~GHz, slightly larger than our working frequency.
When $\theta=90$\textdegree, the ground state transition is on the
contrary at the highest field ($\alpha$ in Fig.~\ref{spec5p5K} and
\ref{vsT}). For an intermediate angle, the transitions can also be
determined by a full diagonalization of the Hamiltonian
(Eq.~\ref{ham}). As a result of the rapid variation of the splittings
with the amplitude and the direction of the applied field, certain
subpopulations of molecules are selected, and well defined absorption
peaks are observed. Fig.~\ref{angle}b, which shows the orientations
contributing to peaks $\alpha$ and $\epsilon$, validates this picture.
The influence of the $C$--term can be seen for the peak $\alpha$
(variation between $\phi=0$\textdegree~and $\phi=45$\textdegree), for
which the easy axis of the contributing molecules almost lie in the
plane $\theta=90$\textdegree~(the band near $\theta=0$\textdegree~is a
highly excited transition which does not contribute at low
temperature).

Finally, echo experiments ($\pi/2-\tau-\pi-\tau-$echo high power pulse
sequences) were conducted at 5.5~K on several dilutions of Ni$_4$
(down to 0.5 mg/mL) and on the pure solvent. We did not observe any
echo from the Ni$_4$ molecules. Let us briefly discuss this result.
First, the sensitivity of the experimental setup is sufficient to
detect echoes with $T_2>50$~ns in 200~G broad lines from micromolar
solutions (\emph{i.e.} $10^{11}$ spins) of $S=1/2$ compounds. Despite
the random orientation of the Ni$_4$ molecules and the broad
absorption lines ($\approx$2 to 4~kG), the cw EPR data shows that a
small ($\approx$5\%) but non negligible fraction of the molecules
contribute to the signal. Taking into account these factors, it is
found that there are still about 30 times more spins than needed to
detect an echo if $T_2>50$~ns for the 10~mg/mL Ni$_4$ solution.
Second, the random orientation makes the intermolecular interaction
weak because the effective dilution of the molecules adressed by the
microwaves is large. Third, the coupling to nuclear spins can yield
coherence times larger than 50~ns \cite{ardavan07}. It is not clear
whether the coherence time of spin-states oscillations between excited
levels (as $\beta$ to $\zeta$) could be much shorter than the one
between the ground state and the first excited states (as $\alpha$).
However, an estimation of the spin-lattice relaxation time from direct
spin-phonon process \cite{chudnovsky05} using a reasonable value for
the transverse speed of sound in the frozen solvent (1000~m$/$s)
yields $T_1=40$~ns at $f=130$~GHz and 5.5~K (it increases by a factor
2 at 0~K). This upper bound for $T_2$ could explain the absence of
echo from Ni$_4$ in our experiments. It is interesting to note that
$T_1$ is a strong function of frequency ($\sim 1/f^3$). For instance,
at 10~GHz and 1.5~K we estimate $T_1=60~\mu$s.

In conclusion, we have studied dilute frozen solutions of the SMM
Ni$_4$ by HF EPR. The clusters have been found to be stable in
solution. Despite the random orientation of the anisotropy axes, it is
possible to measure EPR spectra with well defined transitions and to
deduce that the ZFS parameters of the dissolved molecules are close to
the ones in the crystal phase. Thus, the behavior of an isolated
Ni$_4$ cluster is essentially the same as that of a single crystal.

This work was suppported by NSF-NIRT Grant No. DMR-0506946 (ADK) and
by NIH Grant No. GM075920 (GJG).

\end{document}